\newcommand \Mpc {h^{-1}{\rm Mpc}}
\newcommand \kpc {h^{-1}{\rm kpc}}
\newcommand \farcs{\hbox{$.\!\!^{\prime\prime}$}}

\newcommand \mic {$\mu$m}
\newcommand \kms {{\rm km~s}^{-1}}
\newcommand \msun {h^{-1} M_\odot}
\newcommand \beqn {\begin{equation}}
\newcommand \eeqn {\end{equation}}
\newcommand \Spitzer {{\em Spitzer}~}
\newcommand \Chandra {{\em Chandra}~}
\newcommand \ROSAT {{\em ROSAT}~}
\documentclass[apjl]{emulateapj}

\begin{document}

\title{An Extremely Massive Dry Galaxy Merger in a Moderate Redshift Cluster}

\author{Kenneth Rines\altaffilmark{1,2}, Rose Finn\altaffilmark{3}, and
Alexey Vikhlinin\altaffilmark{1,4}} 
\email{krines@astro.yale.edu}

\altaffiltext{1}{Harvard-Smithsonian Center for Astrophysics, Cambridge, MA 02138; krines@cfa.harvard.edu}
\altaffiltext{2}{Yale Center for Astronomy and Astrophysics, Yale University, New Haven, CT 06520-8121}
\altaffiltext{3}{Department of Physics, Siena College, Loudonville, NY 12211}
\altaffiltext{4}{Space Research Institute, Profsoyuznaya 84/32, Moscow, Russia}

\begin{abstract}

We have identified perhaps the largest major galaxy merger ever seen.
While analysing \Spitzer IRAC images of CL0958+4702, an X-ray-selected
cluster at $z$=0.39, we discovered an unusual plume of stars extending
$\gtrsim$110$\kpc$ outward from the bright central galaxy (BCG).
Three galaxies 1-1.5 mag fainter than the BCG lie within 17 kpc
(projected) of the BCG and are probably participating in the merger.
The plume is detected in all four IRAC channels and at optical
wavelengths in images from the WIYN telescope; the surface brightness
is remarkably high ($\mu_r\approx$24.8 mag arcsec$^{-2}$ at 50
kpc). The optical and infrared colors are consistent with those of
other BCGs, suggesting that the plume is composed of old stars and
negligible recent star formation (hence a ``dry merger'').  The
luminosity in the plume is at least equivalent to a 4$L^*$ galaxy.  A
diffuse halo extending 110$\kpc$ from the BCG in one 
IRAC image suggests the total amount of diffuse light is
$L_r\sim$1.3$\times$10$^{11}h^{-2} L_\odot$.  A \Chandra observation
shows an X-ray image and spectrum typical of moderate-mass clusters.
We use MMT/Hectospec to measure 905 redshifts in a 1 $deg^2$ region
around the cluster.  The velocities of two of the BCG companions
indicate a merger timescale for the companion galaxies of $\sim$110
Myr and $\sim$0.5-1 Gyr for the plume.  We conclude that the BCG and
intracluster light of CL0958 is formed by major mergers at moderate
redshifts.  After the major merger is complete, CL0958 will likely
become a fossil cluster.

\end{abstract}


\section{Introduction}

Galaxies in the centers of clusters likely experience many mergers
during their formation.  Numerical simulations of cluster formation
often show frequent mergers of small galaxies with a larger, central
galaxy \citep[e.g.,][]{dubinski98}.  Supporting this view, many
central cluster galaxies contain multiple components, and some cluster
galaxies at high redshift are apparently merging pairs
\citep{lauer88,vandokkum99,yamada02,tran05}.  One bright galaxy (the 
``Spiderweb'') in a protocluster at $z$=2.16 is apparently undergoing
several minor mergers \citep{miley06}.  Some of these mergers likely
produce the intracluster light seen in nearby clusters
\citep{zwicky1951,2000ApJ...536..561G,mihos05,krick07}.    
Deep optical observations of early-type field galaxies
also reveal features associated with either
ongoing mergers or merger remnants \citep{vandokkum05}.  These
features typically contribute $\sim$5\% of the total starlight:
features of this brightness can be produced by either major or minor
mergers \citep{vandokkum05,kawata06}.  

We report here the biggest major galaxy merger ever seen (to our
knowledge), involving the near-complete disruption of a galaxy
comparable to a brightest cluster galaxy (BCG).  This cosmic collision
is set in CL0958+4702, an X-ray cluster at moderate redshift
($z$=0.39) discovered with {\em ROSAT}.  \citet{molthagen97}
identified a galaxy with $z$=0.39 as the optical counterpart (although
they did not classify it as a cluster).  CL0958 was reidentified in
the 400 Square Degree serendipitous survey (400d) of clusters and groups in
pointed \ROSAT observations \citep{burenin07}.

As part of a multiwavelength observing campaign of 400d clusters, we
discovered that CL0958 displays an unusual plume of emission in
\Spitzer IRAC images.  Further analysis indicates that the emission is
coming from an old stellar population at the redshift of the cluster.
This evidence suggests that CL0958 is undergoing a major dry merger.
The amount of starlight in the plume is at least equal to a 4$L^*$
galaxy.

CL0958 presents a dramatic example of intracluster light; extended
emission is clearly detected in optical images from the WIYN telescope
and in mid-infrared images from \Spitzer IRAC data.  We determine that
CL0958 will probably evolve into a fossil cluster, i.e., a cluster in
which the second-brightest member is $\gtrsim$2 mag fainter
than the BCG \citep{ponman94,jones00}.  Fossil groups and fossil
clusters are thought to be produced either from unusual initial
conditions or from the mergers of several bright cluster members into
one dominant BCG \citep[e.g.,][]{milosavljevic06}.  CL0958 provides a
dramatic example of the merger scenario in progress.

A more complete description of the properties of CL0958 will be made
in \citet{c3poi}.  Here we describe the basic features
of the multiwavelength data.  We assume cosmological parameters of
$\Omega_m$=0.3, $\Omega_\Lambda$=0.7, and $H_0$=100$h \kms Mpc^{-1}$.  At
the redshift of the cluster ($z$=0.39), the spatial scale is
1$''$=3.7 $\kpc$.

\section{Observations}
\subsection{Optical Imaging}

We observed CL0958 (and other 400d clusters) with the WIYN telescope
in 2005 December.  We obtained $griz$ imaging with the OPTIC camera
\citep{tonry02,howell03} with the goal of identifying cluster members
up to 3 mag fainter than $M^*$.  Figure \ref{image} shows the
$r$ image from a non-photometric night with $0\farcs5$ seeing.
Surrounding the main component of the BCG are three faint components
and an asymmetric plume of low surface brightness emission extending
primarily NW of the BCG (outlined by contours).  The morphology of the
plume is very similar to that seen in the late stages of simulations
of dry mergers \citep[e.g., Figure 4 of][Figure 1 of van Dokkum
2005]{dubinski98}.  The extended emission shows little small-scale
structure at the resolution of these images.  Note that merger
features identified in early-type field galaxies similarly show little
small-scale structure in contrast to tidal features seen around
late-type galaxies \citep{vandokkum05}.  Using the Sloan Digital Sky
Survey \citep[SDSS, ][]{sdss} for calibration, we estimate that the
outer contour in Figure \ref{image} is $\mu_r$$\approx$24.8 mag arcsec$^{-2}$.
Corrected for cosmological surface brightness dimming, the plume in
CL0958 has much higher surface brightness ($\mu_r\approx$23.4) than
intracluster light and even some cD envelopes in nearby clusters
\citep[e.g.,][]{schombert88,2000ApJ...536..561G,mihos05,krick07}.

The three companions to the BCG are likely cluster galaxies
participating in the major merger.  Assigning flux to the various
components is challenging.  We use SExtractor \citep{sex} to measure
relative magnitudes of the BCG and its three companions within 1''
diameter apertures: the N,E, and SE companions are respectively
+1.32,+1.09, and +1.29 mag fainter than the BCG.  The BCG has
absolute magnitude $M_r\approx$-21.6 or $L_r\sim$3$\times$10$^{10}
h^{-2} L_\odot$ \citep[we use Table A1 of][to correct for
bandpass shifting and passive evolution]{wake06}.  The plume has a
total magnitude r$\sim$18.8 ($L_r\sim$5$\times$10$^{10}
h^{-2} L_\odot$) and contains $\sim$46\% of the
total light in the region enclosed by the contours in Figure
\ref{image}, although this fraction depends sensitively on the
parameters used to estimate the contribution of the BCG and BCG
companions \citep[e.g., see][]{gonzalez05}.  The plume is visible to a
similar spatial extent in $i$ band, and is marginally detectable in
$g$ and $z$, consistent with the plume having colors typical of an old
stellar population.


\subsection{Mid-Infrared Imaging}

We are observing several 400d clusters with \Spitzer to
measure the star formation rates in X-ray clusters and their
dependence on redshift and cluster mass (GO2 Program 20225).
CL0958 was imaged with IRAC with 100 s exposures using the 12-point
Realeaux dither pattern with ``medium'' dither size.  We processed the
Basic Calibration Data (BCD) files with the Post-BCD software MOPEX
after using the IRAC artifact mitigation codes provided by Sean Carey.

\begin{figure}
\figurenum{2}
\plotone{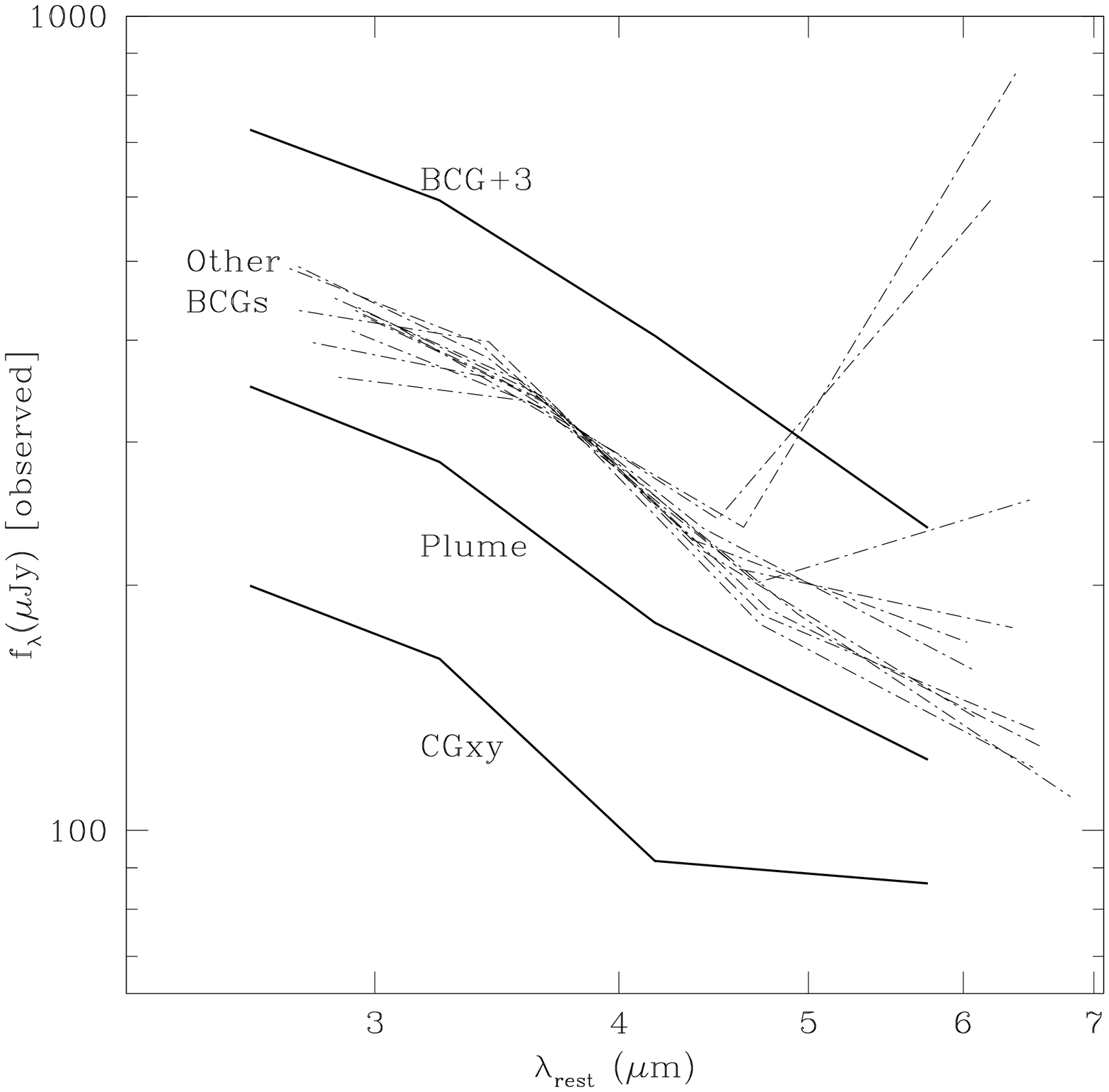}
\caption{\label{sed} 
Spectral energy distribution of the BCG (including companions) and the
plume of emission. SEDs of another cluster member (labeled ``CGxy'')
and of several BCGs from \citet[][dash-dotted lines, scaled to cross
at around 4\mic]{egami06} are shown for
comparison.  Note that the IRAC resolution complicates separating the
BCG light from the companions.  The spectra of the BCG and two of the
companions show early-type populations, so the color differences are
probably small. }
\end{figure}
\epsscale{1}

\begin{figure}
\figurenum{1}
\label{image}
\plottwo{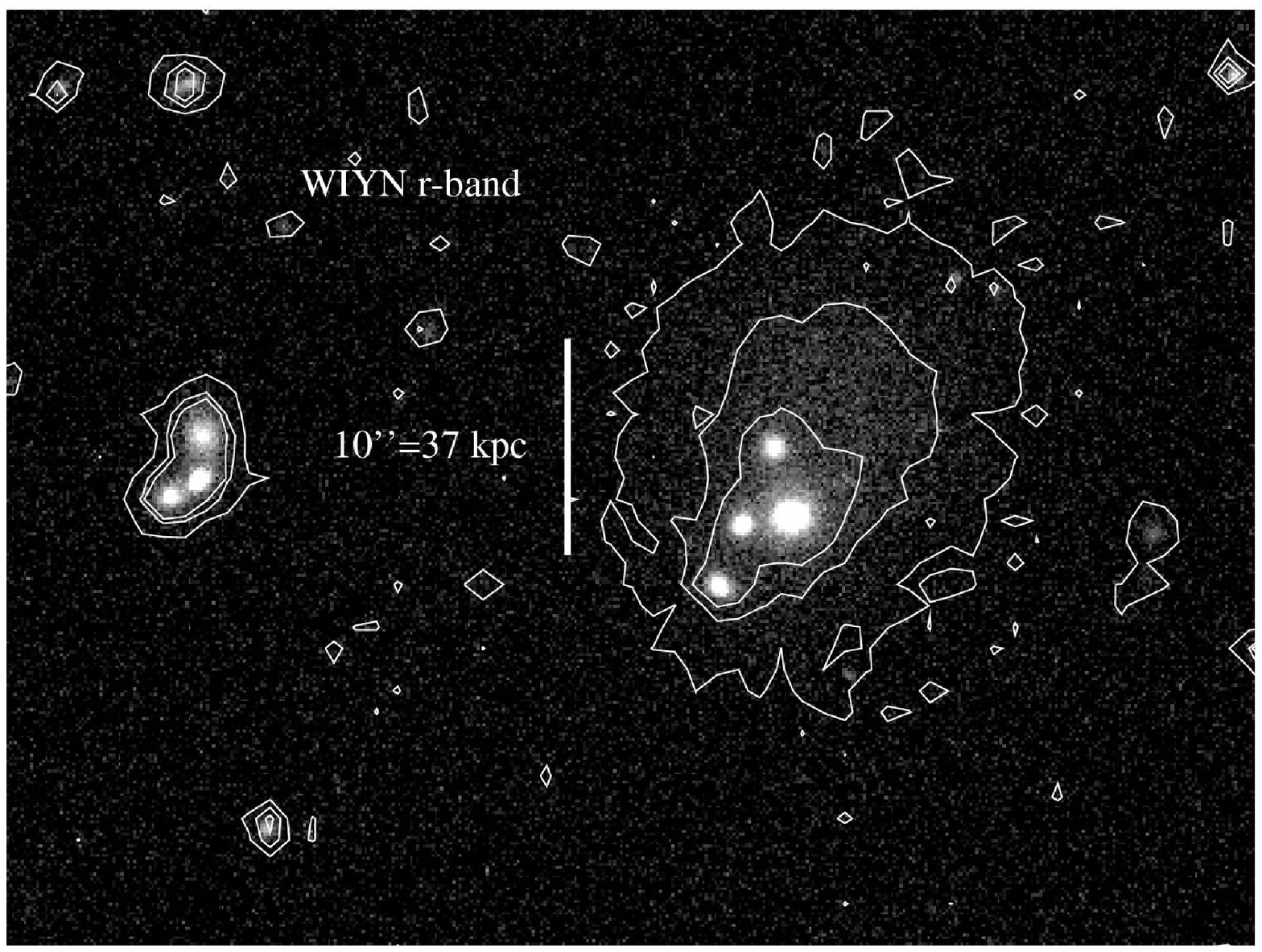}{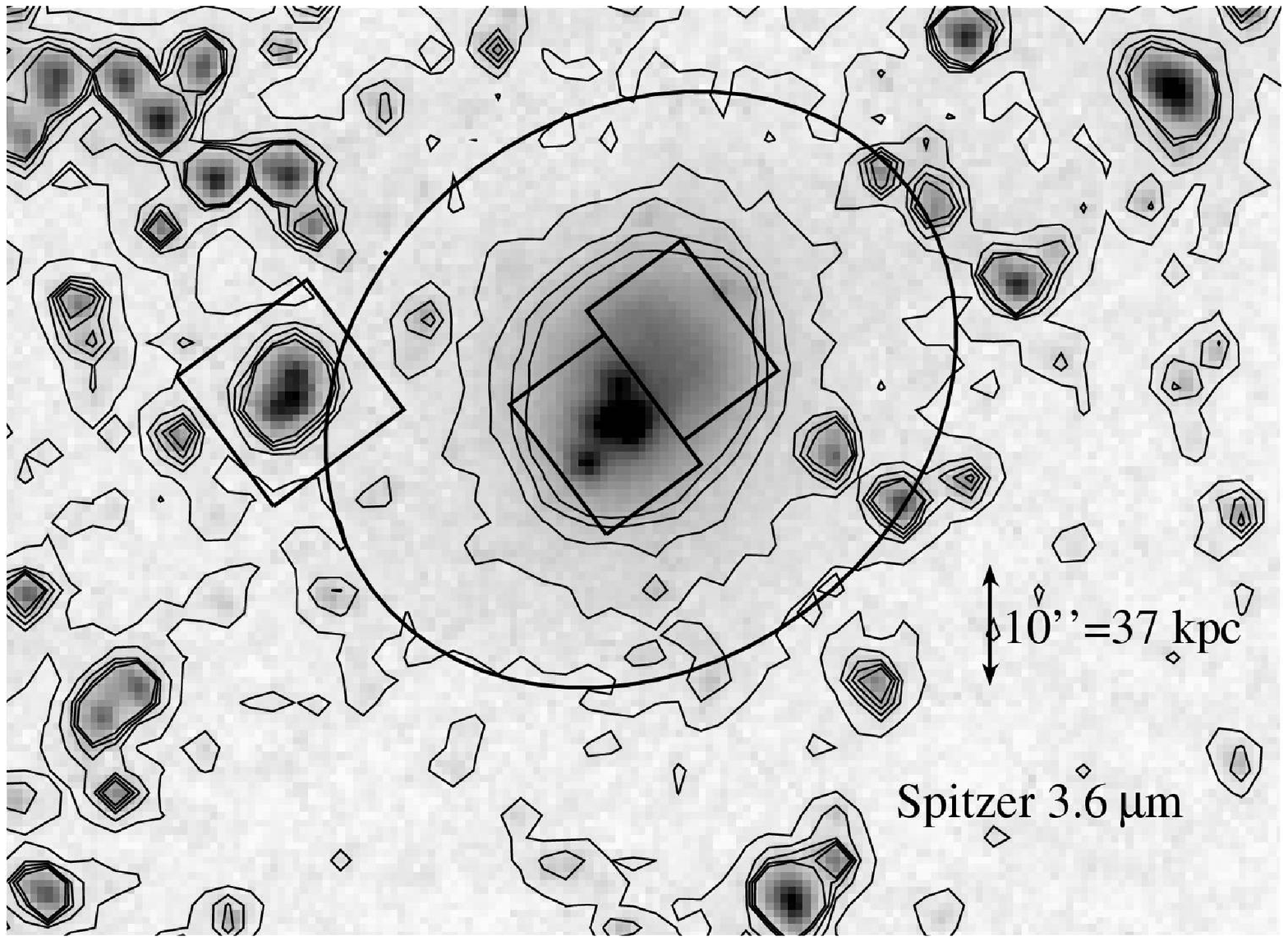} 
\plotone{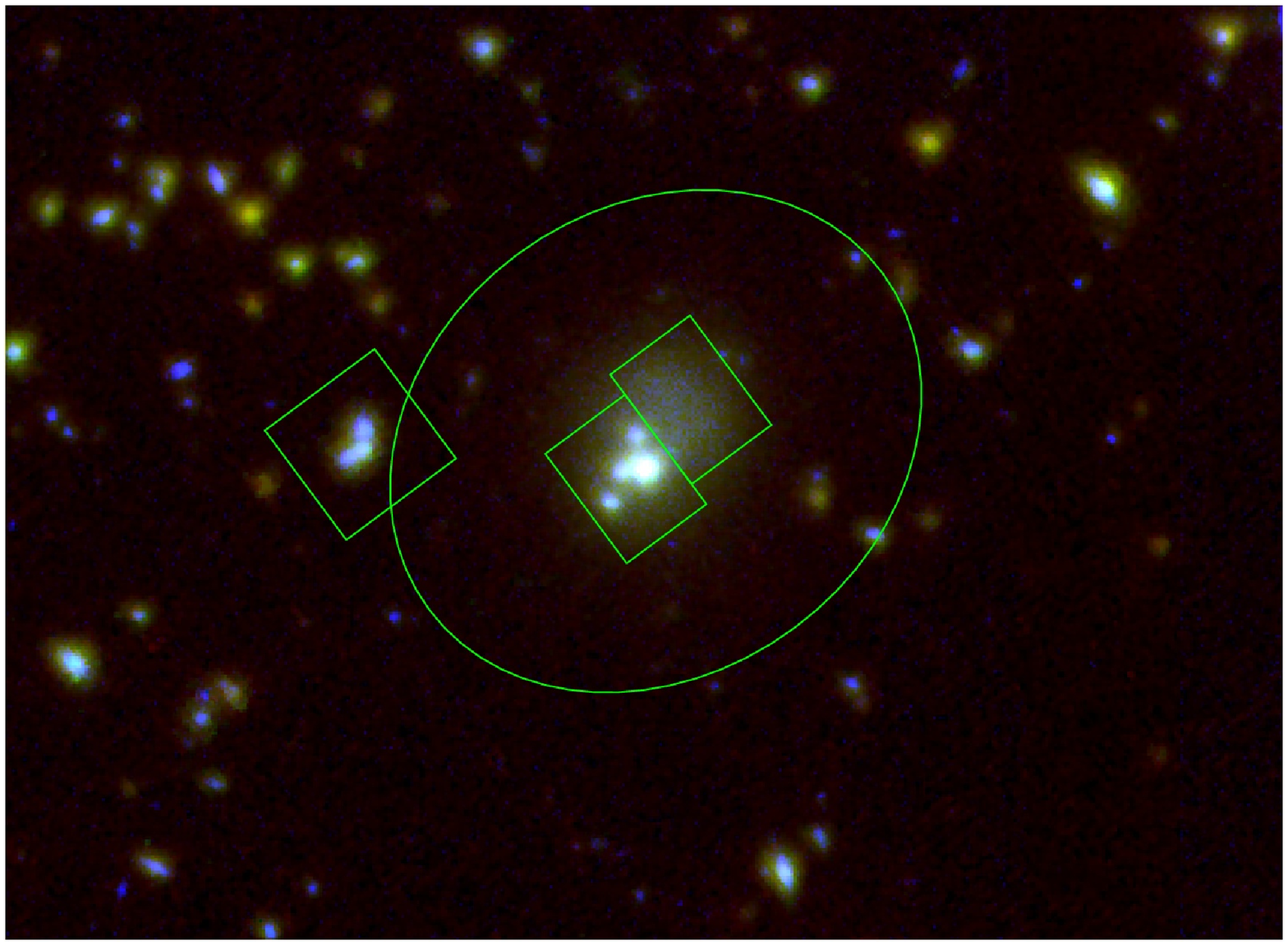}
\caption{Upper left: 2$\times$120-s $r$-band WIYN image taken with OPTIC in 0\farcs5 seeing.  
The contours show surface brightness levels of $\mu_r$ = 24.82,24.79,
and 24.76 mag arcsec$^{-2}$.  The BCG has three companion galaxies
within 17$\kpc$ to the N, E, and SE.  Upper right: IRAC 3.6\mic ~image.
Contours are linearly spaced between 0.07 and 0.11 MJy ster$^{-1}$.
The ellipse approximately indicates the spatial extent of the
plume.  Boxes identify the regions used to measure the SED of the plume, BCG+3, and a nearby cluster galaxy (itself a likely merger in progress).  
Bottom: False-color image of CL0958.  WIYN r-band image in blue, \Spitzer IRAC
3.6\mic ~image in green, IRAC 4.5\mic ~image in red.  All images are oriented with North up and East to the left.
}
\end{figure}

Figure \ref{image} shows IRAC images from the 3.6 and 4.5\mic~
channels.  The emission seen in the WIYN images is clearly visible.
The emission is weakest in the 8.0~\mic~ band, so we measure the colors
of the BCG and the plume in two regions where the emission is
significant at 8.0~\mic.  Specifically, we measure the flux in two
boxes of 11$\times$8 IRAC pixels, or 35$\times$50~$\kpc$ (shown in
Figure \ref{image}) as well as for another cluster galaxy and
background regions.  The box containing the BCG also contains the
three BCG companions.  The IRAC colors (Figure \ref{sed}) agree very well with those
of non-star-forming BCGs observed by \citet[][]{egami06}.  
The luminosity contained in the box
around the plume is $\approx$48\% of that contained in the box
containing the BCG and companions.  Note that the luminosity of the
plume may be slightly overestimated due to both light from the wings
of the PSF of the BCG and to the IRAC calibration technique, which is
optimized for photometry of point
sources.\footnote{See http://ssc.spitzer.caltech.edu/irac/calib/extcal/index.html}
These two effects could lead to a $\sim$10-20\% overestimate of the
flux of the plume within this box.  Intracluster light and merger
remnants in the nearby universe typically extend to extremely low
surface brightnesses
\citep[e.g.,][]{mihos05,vandokkum05}, so the above is probably a lower
limit on the total light in the plume.  Indeed, the 3.6~\mic ~image
clearly shows extended low surface brightness emission
$\gtrsim$110$\kpc$ from the BCG (Fig.~\ref{image}).  Using
SExtractor to estimate the light in different apertures, we estimate
the total diffuse light in this halo to be $L_r\sim$1.3$\times 10^{11}
h^{-2} L_\odot$ [rest-frame $r$-band corrected for passive evolution;
we estimate $(r-3.6\mu m)$$\approx$4.2 from 8 spectroscopically confirmed
cluster members].  Comparison to the BCGs observed by
\citet{egami06} shows that CL0958 is anomalous: most moderate-redshift
clusters contain several bright galaxies near the X-ray center, and
none show extended emission similar to that in CL0958.


The IRAC colors of both the BCG and the plume indicate that the
ongoing merger is a dry merger: i.e., there is little or no star
formation triggered by the merger.  This result shows that the
progenitor galaxies contained little or no gas, suggesting that the
gas around cluster galaxies is efficiently removed by the cluster,
e.g., by tidal stripping or ram pressure stripping by the intracluster
medium \citep{gunngott,merritt85}.  The fact that the plume is
composed of an old stellar population demonstrates that the plume is
not created by recent star formation, e.g., from a massive cooling
flow not centered on the BCG.  Further, comparison to the BCGs of
\citet{egami06} indicates that the plume has less recent star
formation than some BCGs.  We conclude that the plume was recently
created through a ``dry'' merger, i.e., a merger where the progenitor
galaxies contain little or no gas such that little star formation is
triggered by the merger \citep[e.g.,][]{bell06}.

\subsection{X-ray Observations}

CL0958 was observed with \Chandra for 35 ks as part of a \Chandra survey of 400d clusters.
Complete details on these observations will be presented in
\citet{cccp}.  The X-ray emission is regular and shows a slight
elongation in the direction of the stellar plume.  The spectrum yields
a temperature of $3.6\pm0.8$keV.  The X-ray luminosity is
$L_X$=5.0$\times$10$^{43} h^{-2}$erg s$^{-1}$, making CL0958 a
moderately luminous cluster, intermediate in both $L_X$ and $T_X$ to
Virgo and Abell 2199 \citep{cirsi}.  There is no evidence for a
cooling flow in CL0958, supporting our conclusion from the IRAC colors
that the plume is not created by cooling of the ICM and subsequent
star formation.  The X-ray data show little evidence of the dynamical
activity suggested by the extended optical light.  We find no evidence
of any AGNs in the \Chandra field.  The upper limit on AGNs at the
cluster redshift is $L_X$$<$6$\times 10^{41}$erg s$^{-1}$.

\subsection{Optical Spectroscopy}

We obtained optical spectroscopy of CL0958 with MMT/Hectospec with the
270-line grating \citep[yielding 6.2\AA~FWHM resolution,
][]{hectospec} in 2007 February.  Because the Hectospec field
(1$^\circ$ diameter) is much larger than the WIYN images
($10'$$\times$$10'$), we used imaging data from SDSS to select targets
for spectroscopy.  We observed CL0958 with four configurations (one in
marginal conditions) and obtained 905 redshifts.  A full discussion of
the spectroscopy will be presented in \citet{c3poi}.  Here we
highlight the results relevant to the major merger.

A spectrum of the BCG shows a typical early-type galaxy with no
evidence of recent star formation from either [OII] in emission or
H$\delta$ in absorption.  The 4000\AA~break is strong, typical of an
old stellar population.  We obtained spectra of the N and SE BCG
companions, which have early-type spectra and rest-frame velocities of
-63$\pm$41$\kms$ and +86$\pm$41$\kms$ respectively relative to the BCG
\citep[redshift uncertainties estimated from repeat observations, 
see][]{hectospec}.  These redshifts provide strong evidence that the N
and SE BCG companions are physically close to the BCG; it is highly
unlikely that they are chance projections of cluster members from the
outskirts of the cluster.  This close association supports the
possibility that these companions are the remaining nuclei of larger
galaxies disrupted in the merger.

We identify 21 galaxies within 1 Abell radius of the BCG as likely
cluster members based on the redshift-radius distribution
\citep{c3poi}.  These members yield a projected velocity dispersion of
$\sigma_p$=521$^{+107}_{-66}\kms$ using the procedure of
\citet{danese}.  However, there is a noticeable core of 11 galaxies
within 0.8$\Mpc$ with a projected velocity dispersion of
$\sigma_p$=244$^{+80}_{-41}\kms$.  Assuming the galaxies trace the
dark matter, the scaling relation of \citet{evrard07} yields a virial
mass estimate of $M_{200}\approx$9.2 (1.0)$\times$10$^{13}\msun$ using
the upper (lower) estimates of the velocity dispersion.
Identification of more cluster members is
required to robustly estimate the projected velocity dispersion.

\section{Discussion}

With the multiwavelength data, we can estimate the timescale of the
merger.  The spatial scale is set by the limit of the plume, which
extends to $\approx$110~$kpc$ in the 3.6~\mic ~image.  Assuming that the
plume has similar velocity to the BCG companions, the velocity scale
is $\lesssim$100$\kms$.  These scales suggest a timescale of order
0.5-1 Gyr for the merger remnant to form a relaxed distribution around
the BCG.  This timescale suggests that major mergers might be fairly
common in deep images of galaxy clusters.

The BCG companions lie much closer to the BCG than the plume; thus,
they will likely merge more quickly.  The dynamical friction timescale
can be estimated as
\beqn
T_{\mbox{fric}} = \frac{2\times10^5 r^2 v_c}{M \log{\Lambda}}\mbox{Gyr}
\eeqn
\citep{patton00}.  
Correcting for projection assuming random orientation of $r_p$ and
$v_p$ [$r^2v_c = (3/2)r_p^2(\sqrt{3})v_p$], we estimate
$T_{\mbox{fric}}$$\approx$110 Myr for a companion galaxy mass of
$\approx5\times10^{10}M_\odot$ (the BCG companions are $\approx$$L^*$
galaxies and we assume $M/L_r\approx 5 M_\odot/L_\odot$).  We are
therefore observing CL0958 at a fairly special time in its history.
The unusual properties of CL0958 have implications for at least two
categories of cluster studies: intracluster light and the origin of fossil groups and clusters.

\subsection{Intracluster Light}

The unusual extended emission around the BCG of CL0958 provides a
powerful confirmation of the importance of intracluster light to the
total optical luminosity and direct evidence for its origin.  The
surface brightness and spatial extent of the emission is perhaps the
largest detected to date.  The dearth of cluster galaxies (other than
the BCG companions) with magnitudes comparable to the BCG suggests
that the BCG is forming via a major merger.  The intracluster light
could be the leftover stars from bright galaxies that have recently
been disrupted or stars that have been stripped from the BCG
companions.  CL0958 demonstrates that at least some intracluster light
is produced by extreme dynamic events in cluster centers as opposed to
slower mechanisms.  Indeed, the detailed simulations of
\citet{murante07} suggest that most intracluster light is produced by
the mergers that form the BCG rather than from tidal stripping.

Major ongoing dry mergers appear to be rare in BCGs: images of 35 BCGs
at $z$$<$0.1 in Figure 3 of \citet{vonderlinden07} show no mergers
similar to that in CL0958, although several BCGs have multiple
components: these may be major mergers observed at an earlier stage
\citep{lauer88}.  CL0958 indicates that ongoing mergers may be more
common at higher redshift, as would be expected in hierarchical models
of BCG formation \citep{delucia07}.  

\subsection{The Origin of ``Fossil'' Groups and Clusters}

The multiwavelength properties of CL0958 clearly indicate that it is a
moderately massive cluster with X-ray properties (and possibly also
velocity dispersion) comparable to (1-5)$\times$$10^{14}\msun$
clusters.  Once the three BCG companions merge with the BCG, the BCG
will be $\sim$1.3 mag brighter than the next-brightest galaxy (in
observed $r$-band).  Depending on the fraction of the plume that
accretes onto the BCG, within $\sim$1 Gyr the BCG will be $\sim$2 mag
brighter than the next-brightest galaxy.  These properties would make
CL0958 a ``fossil cluster'', that is, a fossil group with
$M\gtrsim10^{14}\msun$
\citep{khosroshahi06}.

There are two basic mechanisms that could produce fossil groups and
fossil clusters.  Either initial conditions of the forming group lead
to the formation of an extremely dominant BCG/BGG, or several bright
group galaxies merge after the initial collapse of the group.  One
difficulty with the merger hypothesis is that dynamical friction is
less efficient for cluster galaxies that have been tidally stripped by
the cluster potential \citep{merritt85}.  However, more recent
theoretical models suggest that cluster merger variance should
occasionally produce such systems \citep{milosavljevic06}.
The major dry merger in CL0958 shows that fossil clusters and
groups can be produced by major mergers of bright galaxies in the
system centers.  Deep optical and infrared observations of the other known
fossil clusters and groups may reveal less dramatic evidence of merger
activity.

\section{Conclusions}

We have identified a massive stellar plume formed by a major dry
merger in an X-ray cluster at $z$=0.39.  The amount of light contained
in the merger suggests that the equivalent of a 4$L^*$ galaxy in stars
is currently visible as a merger remnant; an additional
$\sim$10$^{11}L_\odot$ is visible in a diffuse halo of intracluster
light.  To our knowledge, this is the most massive major galaxy merger
ever identified \citep[the ``Spiderweb'' galaxy is an ensemble of
minor mergers][]{miley06}.  The colors of the plume indicate that it
is composed of an old stellar population, consistent with origin in
early-type cluster galaxies and inconsistent with recent star
formation.  This remnant will likely merge with the BCG in $\lesssim$1
Gyr, while three galaxies very close to the BCG will probably merge
within $\sim$110 Myr.  After these galaxies and the plume have merged
with the BCG, CL0958 will likely form a ``fossil cluster''.
Theoretical models of BCG assembly indicate that BCGs can be assembled
quite late, but these models predict that most of this growth occurs
through minor mergers
\citet{delucia07}.  Possible substructure in the galaxy dynamics
complicates the measurement of the velocity dispersion.
The existence of this major dry merger demonstrates that
intracluster light can be formed in extreme dynamic events and that
fossil clusters are at least sometimes formed by major mergers in
their centers.

\acknowledgements

We thank the referee for suggestions that improved the presentation of
 this letter.
We thank Margaret Geller, Michael Kurtz, Scott Kenyon, Sune Toft,
 Rodion Burenin, Thomas Reiprich, Harald Ebeling, Allan Hornstrup, and
 Hernan Quintana for helpful discussions and suggestions.
 Observations reported here were obtained at the MMT Observatory, a
 joint facility of the Smithsonian Institution and the University of
 Arizona.  
The WIYN Observatory is a joint
 facility of the University of Wisconsin-Madison, Indiana University,
 Yale University, and the National Optical Astronomy Observatories.
This work is based on observations made with the Spitzer Space Telescope, which is operated by the Jet Propulsion Laboratory, California Institute of Technology under a contract with NASA. Support for this work was provided by NASA through an award issued by JPL/Caltech.


\begin{thebibliography}{39}
\expandafter\ifx\csname natexlab\endcsname\relax\def\natexlab#1{#1}\fi

\bibitem[{{Bell} {et~al.}(2006)}]{bell06}
{Bell}, E.~F. {et~al.} 2006, \apj, 640, 241

\bibitem[{{Bertin} \& {Arnouts}(1996)}]{sex}
{Bertin}, E. \& {Arnouts}, S. 1996, \aaps, 117, 393

\bibitem[{{Burenin} {et~al.}(2006){Burenin}, {Vikhlinin}, {Hornstrup},
  {Ebeling}, {Quintana}, \& {Mescheryakov}}]{burenin07}
{Burenin}, R.~A., {Vikhlinin}, A., {Hornstrup}, A., {Ebeling}, H., {Quintana},
  H., \& {Mescheryakov}, A. 2006, (astro-ph/0610739)

\bibitem[{{Danese} {et~al.}(1980){Danese}, {de Zotti}, \& {di Tullio}}]{danese}
{Danese}, L., {de Zotti}, G., \& {di Tullio}, G. 1980, \aap, 82, 322


\bibitem[{{De Lucia} \& {Blaizot}(2007)}]{delucia07}
{De Lucia}, G. \& {Blaizot}, J. 2007, \mnras, 375, 2

\bibitem[{{Dubinski}(1998)}]{dubinski98}
{Dubinski}, J. 1998, \apj, 502, 141

\bibitem[{{Egami} {et~al.}(2006)}]{egami06}
{Egami}, E. {et~al.} 2006, \apj, 647, 922

\bibitem[{{Evrard} {et~al.}(2007)}]{evrard07}
{Evrard}, A.~E. {et~al.} 2007, ApJ, submitted (astro-ph/0702241)

\bibitem[{{Fabricant} {et~al.}(2005)}]{hectospec}
{Fabricant}, D. {et~al.} 2005, \pasp, 117, 1411

\bibitem[{{Gonzalez} {et~al.}(2005){Gonzalez}, {Zabludoff}, \&
  {Zaritsky}}]{gonzalez05}
{Gonzalez}, A.~H., {Zabludoff}, A.~I., \& {Zaritsky}, D. 2005, \apj, 618, 195

\bibitem[{{Gonzalez} {et~al.}(2000){Gonzalez}, {Zabludoff}, {Zaritsky}, \&
  {Dalcanton}}]{2000ApJ...536..561G}
{Gonzalez}, A.~H., {Zabludoff}, A.~I., {Zaritsky}, D., \& {Dalcanton}, J.~J.
  2000, \apj, 536, 561

\bibitem[{{Gunn} \& {Gott}(1972)}]{gunngott}
{Gunn}, J.~E. \& {Gott}, J.~R.~I. 1972, \apj, 176, 1

\bibitem[{{Howell} {et~al.}(2003){Howell}, {Everett}, {Tonry}, {Pickles}, \&
  {Dain}}]{howell03}
{Howell}, S.~B., {Everett}, M.~E., {Tonry}, J.~L., {Pickles}, A., \& {Dain}, C.
  2003, \pasp, 115, 1340

\bibitem[{{Jones} {et~al.}(2000){Jones}, {Ponman}, \& {Forbes}}]{jones00}
{Jones}, L.~R., {Ponman}, T.~J., \& {Forbes}, D.~A. 2000, \mnras, 312, 139

\bibitem[{{Kawata} {et~al.}(2006){Kawata}, {Mulchaey}, {Gibson}, \&
  {S{\'a}nchez-Bl{\'a}zquez}}]{kawata06}
{Kawata}, D., {Mulchaey}, J.~S., {Gibson}, B.~K., \&
  {S{\'a}nchez-Bl{\'a}zquez}, P. 2006, \apj, 648, 969

\bibitem[{{Khosroshahi} {et~al.}(2006){Khosroshahi}, {Maughan}, {Ponman}, \&
  {Jones}}]{khosroshahi06}
{Khosroshahi}, H.~G., {Maughan}, B.~J., {Ponman}, T.~J., \& {Jones}, L.~R.
  2006, \mnras, 369, 1211

\bibitem[{{Krick} \& {Bernstein}(2007)}]{krick07}
{Krick}, J.~E. \& {Bernstein}, R.~A. 2007, ArXiv e-prints, 704

\bibitem[{{Lauer}(1988)}]{lauer88}
{Lauer}, T.~R. 1988, \apj, 325, 49

\bibitem[{{Merritt}(1985)}]{merritt85}
{Merritt}, D. 1985, \apj, 289, 18

\bibitem[{{Mihos} {et~al.}(2005){Mihos}, {Harding}, {Feldmeier}, \&
  {Morrison}}]{mihos05}
{Mihos}, J.~C., {Harding}, P., {Feldmeier}, J., \& {Morrison}, H. 2005, \apjl,
  631, L41

\bibitem[{{Miley} {et~al.}(2006)}]{miley06}
{Miley}, G.~K. {et~al.} 2006, \apjl, 650, L29

\bibitem[{{Milosavljevi{\'c}} {et~al.}(2006){Milosavljevi{\'c}}, {Miller},
  {Furlanetto}, \& {Cooray}}]{milosavljevic06}
{Milosavljevi{\'c}}, M., {Miller}, C.~J., {Furlanetto}, S.~R., \& {Cooray}, A.
  2006, \apjl, 637, L9

\bibitem[{{Molthagen} {et~al.}(1997){Molthagen}, {Wendker}, \&
  {Briel}}]{molthagen97}
{Molthagen}, K., {Wendker}, H.~J., \& {Briel}, U.~G. 1997, \aaps, 126, 509

\bibitem[{{Murante} {et~al.}(2006)}]{murante07}
{{Murante}, G. and {Giovalli}, M. and {Gerhard}, O. and {Arnaboldi}, M. and 
	{Borgani}, S. and {Dolag}, K.}  2007, \mnras, 377, 2

\bibitem[{{Patton} {et~al.}(2000){Patton}, {Carlberg}, {Marzke}, {Pritchet},
  {da Costa}, \& {Pellegrini}}]{patton00}
{Patton}, D.~R., {Carlberg}, R.~G., {Marzke}, R.~O., {Pritchet}, C.~J., {da
  Costa}, L.~N., \& {Pellegrini}, P.~S. 2000, \apj, 536, 153

\bibitem[{{Ponman} {et~al.}(1994){Ponman}, {Allan}, {Jones}, {Merrifield},
  {McHardy}, {Lehto}, \& {Luppino}}]{ponman94}
{Ponman}, T.~J., {Allan}, D.~J., {Jones}, L.~R., {Merrifield}, M., {McHardy},
  I.~M., {Lehto}, H.~J., \& {Luppino}, G.~A. 1994, \nat, 369, 462


\bibitem[{{Rines} {et~al.}(2007)}]{c3poi}
{Rines}, K. {et~al.} 2007, in preparation

\bibitem[{{Rines} \& {Diaferio}(2006)}]{cirsi}
{Rines}, K.~J. \& {Diaferio}, A. 2006, \aj, 132, 1275

\bibitem[{{Schombert}(1988)}]{schombert88}
{Schombert}, J.~M. 1988, \apj, 328, 475

\bibitem[{{Stoughton} {et~al.}(2002)}]{sdss}
{Stoughton}, C. {et~al.} 2002, \aj, 123, 485

\bibitem[{{Tonry} {et~al.}(2002){Tonry}, {Luppino}, {Kaiser}, {Burke}, \&
  {Jacoby}}]{tonry02}
{Tonry}, J.~L., {Luppino}, G.~A., {Kaiser}, N., {Burke}, B.~E., \& {Jacoby},
  G.~H. 2002, SPIE, 4836, 206

\bibitem[{{Tran} {et~al.}(2005){Tran}, {van Dokkum}, {Franx}, {Illingworth},
  {Kelson}, \& {Schreiber}}]{tran05}
{Tran}, K.-V.~H., {van Dokkum}, P., {Franx}, M., {Illingworth}, G.~D.,
  {Kelson}, D.~D., \& {Schreiber}, N.~M.~F. 2005, \apjl, 627, L25

\bibitem[{{van Dokkum}(2005)}]{vandokkum05}
{van Dokkum}, P.~G. 2005, \aj, 130, 2647

\bibitem[{{van Dokkum} {et~al.}(1999){van Dokkum}, {Franx}, {Fabricant},
  {Kelson}, \& {Illingworth}}]{vandokkum99}
{van Dokkum}, P.~G., {Franx}, M., {Fabricant}, D., {Kelson}, D.~D., \&
  {Illingworth}, G.~D. 1999, \apjl, 520, L95

\bibitem[{{Vikhlinin} {et~al.}(2007)}]{cccp}
{Vikhlinin}, A. {et~al.} 2007, in preparation

\bibitem[{{von der Linden} {et~al.}(2006){von der Linden}, {Best}, {Kauffmann},
  \& {White}}]{vonderlinden07}
{von der Linden}, A., {Best}, P.~N., {Kauffmann}, G., \& {White}, S.~D.~M.
  2006, MNRAS, in press (astro-ph/0611196)

\bibitem[{{Wake} {et~al.}(2006)}]{wake06}
{Wake}, D.~A. {et~al.} 2006, \mnras, 372, 537


\bibitem[{{Yamada} {et~al.}(2002){Yamada}, {Koyama}, {Nakata}, {Kajisawa},
  {Tanaka}, {Kodama}, {Okamura}, \& {De Propris}}]{yamada02}
{Yamada}, T., {Koyama}, Y., {Nakata}, F., {Kajisawa}, M., {Tanaka}, I.,
  {Kodama}, T., {Okamura}, S., \& {De Propris}, R. 2002, \apjl, 577, L89

\bibitem[{{Zwicky}(1951)}]{zwicky1951}
{Zwicky}, F. 1951, \pasp, 63, 61

\end{thebibliography}

\clearpage

\end{document}